\documentclass[%
 reprint,
 amsmath,amssymb,
 aps,
]{revtex4-1}

\usepackage{graphicx}% Include figure files
\usepackage{dcolumn}% Align table columns on decimal point
\usepackage{bm}% bold math
\usepackage{color}

\newcommand{\Z}{\mathbb Z}

\newcommand{\dif}{\mbox{d}}
\newcommand{\BMSt}{\ensuremath{\mathfrak{bms}_3\ }}

\begin{document}

\preprint{ICCUB 16-041, UTTG-25-16}

\title{Canonical Realization of (2+1)-dimensional Bondi-Metzner-Sachs symmetry}% Force line breaks with \\

\author{Carles Batlle}
\email{carles.batlle@upc.edu}
\affiliation{
Departament de Matem\`atiques and IOC, 
Universitat Polit\`ecnica de Catalunya\\
 EPSEVG, Av. V. Balaguer 1, E-08808 Vilanova i la Geltr\'u, Spain
}%

\author{V\'{i}ctor Campello}
\email{vicmancr@gmail.com}
\author{Joaquim Gomis}%
\email{gomis@ecm.ub.edu}
\altaffiliation[also at ]{Theory Group, Department of Physics, University of Texas  Austin, TX, 78712}
\affiliation{%
Departament de F\'{\i}sica Qu\`antica i Astrof\'{\i}sica and Institut de Ci\`encies del Cosmos\\
Universitat de Barcelona, Mart\'i i Franqu\`es 1, E-08028 Barcelona, Spain
}%

\date{\today}% It is always \today, today,

\begin{abstract}
We construct  canonical realizations of the \BMSt algebra 
as symmetry algebras of a free 
Klein-Gordon (KG) field in $2+1$ dimensions, for both the massive and massless case. 
We consider two types of realizations, one on-shell, written in terms of the Fourier modes of the scalar field, and the  other one off-shell with non-local transformations written in terms of the KG field and its momenta. These realizations contain both supertranslations and superrotations, for which we construct the corresponding Noether charges.
\begin{description}

\item[PACS numbers] 11.30.-j   11.10.-z
 
\end{description}
\end{abstract}

\pacs{Valid PACS appear here}% PACS, the Physics and Astronomy
\maketitle

%\tableofcontents

\section{\label{sec:level1}Introduction}

Recently there has been a renewed interest in the BMS group \cite{BMS-1}. 
The BMS invariance of the  gravitational scattering matrix has been proved in
\cite{Strominger:2013jfa} and, as a consequence of this result, Weinberg's soft graviton theorems \cite{Weinberg:1965nx} can be understood as the Ward identities of 
BMS supertranslations \cite{He:2014laa}\cite{He:2014cra}\cite{Campiglia:2015qka}\cite{Kapec:2015ena}.  The relation between supertranslations,
gravitational memory and soft gravitons theorems has also been studied \cite{Strominger:2014pwa}. There is a proposal that the information paradox \cite{Hawking:1976ra} could be understood in  terms of black hole soft hair associated to supertranslations and superrotations  charges \cite{Hawking:2015qqa} \cite{Hawking:2016msc}. On the other hand,  the BMS group could play a crucial role  in understanding holography in 
asymptotically 
flat space times \cite{Banks:2003vp}  \cite{deBoer:2003vf}  \cite{Arcioni:2003xx} \cite{Barnich:2010eb}. BMS symmetry is an infinite conformal extension of the Carroll symmetry \cite{Duval:2014}, which was introduced in \cite{Levy-Leblond} as a limit of the Poincar\'e algebra when the velocity of light is scaled down to zero. A pedagogical overview of the role of BMS symmetries in most of these topics is presented in \cite{Strominger:2017}.

In this paper we construct a canonical realization of the \BMSt algebra \cite{Ashtekar:1996cd}\cite{Barnich:2006av} with supertranslations and superrotations \cite{Barnich:2010eb} associated to a free 
Klein-Gordon (KG) field in $2+1$ dimensions, for both  massive and massless fields. 
Following the  procedure in \cite{Longhi:1997zt},  we consider, in the massive case, the mass-shell hyperboloid representation of the hyperbolic plane $\text{H}_2$, and compute
the associated Laplace-Beltrami operator. It turns out that the three dimensional momenta is an eigenfunction of the differential operator with  eigenvalue $\frac{2}{m^2}$, where $m$ is the mass of the scalar field and the $2$ comes from  
the dimension of the hyperboloid. This property suggests 
to compute all the eigenfunctions of this operator  corresponding to  that eigenvalue
{ with the same asymptotic properties that the three momenta.}
This  allows to generalize the momenta to an infinite set of supermomenta. These momenta yield an infinite dimensional representation of the (2+1) Lorentz group,
and leads to the definition of the generators of the supertranslations 
in terms of the Fourier modes of  the KG field. 

The mass-shell condition for a massless scalar field results in a cone, for which a Laplace-Beltrami operator cannot be constructed. To get around this, we consider the massless limit of the Laplace-Beltrami operator on the hyperboloid \cite{Gomis:2015ata}. 
Once we have the suitable differential operator, the construction goes in parallel with the massive case. 

We also construct a generalization of the Lorentz generators which corresponds to superrotations. { In the massless case,  the algebra of supertranslations and superrotations is the \BMSt algebra introduced in \cite{Barnich:2011ct}. In the massive case, the superrotation generators that we introduce must be separated into two different sets which both contain  the Lorentz part, and each set corresponds to a subalgebra of \BMSt.}
It should be noted that the differential operators appearing in our construction are one of { the two}   Casimirs of the 2+1 Lorentz group.

At the quantum level, the Hilbert space of one-particle states supports a unitary irreducible representation of the Poincare group, and at the same time a unitary reducible representation of the BMS$_3$ group. In contrast with the gravitational approach, our canonical realization of the supertranslations symmetry is not spontaneously broken. Unitary representations of
BMS$_3$ have been also considered in \cite{Barnich:2014kra}\cite{Barnich:2015uva}.

%Summing up, we obtain  two representations, one for the massive case  and another for the massless one, of the \BMSt algebra \cite{Barnich:2011ct}.
 We study  the off-shell (Noether) supertranslation and superrotation symmetries of the massless Klein-Gordon action, and compute the associated Noether charges. These charges  are expressed as non-local linear functionals of fields and momenta. The same construction is carried out for the supertranslations in the massive case.

The organization of the paper is as follows. In Section 2 we construct the supertranslations and superrotations in terms of the Fourier modes of the KG field. In Section 3 we construct the transformations in terms of fields and momenta. Section 4 is devoted to conclusions and outlook. Appendix A presents explicit forms for some of the functions that appear in the non-local transformations obtained in Section 3, {and Appendix B discusses the geometry of the mass-shell hyperboloid in (2+1) dimensions}. We use the Minkowski metric
 $(-++)$ throughout the paper.

\section{\label{CCSa}Canonical realization of  BMS$_3$}

\subsection{Canonical realization of Poincar\'e symmetry for a scalar field}
%%%%%%%%%%%%%%%%%%%%%%%%%%%%%%%%%%%%%%%%%%%%%%%
The lagrangian density for a real massive scalar field is given by 
\begin{equation}\label{action}
\mathcal{L} = -\dfrac{1}{2}\partial_\mu \phi \partial^\mu \phi - \dfrac{1}{2}m^2 \phi^2.
\end{equation}
 The solution of the Klein-Gordon equation, in terms of Fourier modes $a(\vec{k})$, is 
\begin{equation}\label{eom}
\phi(t,\vec{x})=\int \tilde{dk}\, \left(a(\vec{k}) e^{ikx} + \bar{a}(\vec{k}) e^{-ikx}\right),
\end{equation}
where the phase space Fourier modes have the Poisson bracket
\begin{equation}\label{PBaa}
	\{ a(\vec k) , \bar a (\vec q)  \} = -i \Omega(\vec k) \delta^2(\vec k - \vec q).
\end{equation}
The Lorentz invariant integration measure in the hyperbolic plane $\text{H}_2$ is
\begin{equation}\label{dLinv}
\tilde{\dif k}= \frac{\dif^2 k}{\Omega(\vec k)}, \quad \Omega(\vec k) = (2\pi)^2 2 k^0(\vec k) = (2\pi)^2 2 \sqrt{\vec{k}^2+m^2}.
\end{equation}
Noether's theorem allows us to write down the expression for the conserved charge under translations. By use of the solution of the equations of motion (\ref{eom}) the 
 charges on-shell can be written as 
\begin{equation}\label{momenta}
P^\mu = \int \tilde{dk}\, \bar{a}(\vec{k}) k^\mu a(\vec{k}),
\end{equation}
and their action on the Fourier modes is given by
\begin{equation}\label{momentaFourier}
	\left\{P^\mu , a(\vec k)\right\} = i k^\mu a(\vec k).
\end{equation}

The analogous Lorentz charges on-shell are
\begin{equation}
M^{ij} = -i\int\tilde{dk}\, \bar{a}(\vec{k}) \left(k^i \dfrac{\partial}{\partial k^j} - k^j \dfrac{\partial}{\partial k^i}\right) a(\vec{k})
\end{equation}
for rotations, and
\begin{equation} \label{eq:TrueL}
M^{0j} = t P^j - i \int \tilde{dk}\, \bar{a}(\vec{k}) k^0 \dfrac{\partial}{\partial k^j} a(\vec{k})
\end{equation}
for boosts. We define the truncated time-independent Lorentz generators 
\begin{equation} \label{eq:redefM}
{M'}^{ij} = M^{ij}, \quad {M'}^{0j} = M^{0j} - t P^j
\end{equation}
that satisfy the Poincar\'e algebra as well, and have the following Poisson brackets (we drop the prime and will work with these generators henceforth unless otherwise stated)
\begin{gather}
\{P^\mu, P^\nu\} = 0, \quad \{M^{\mu\nu}, P^\rho\} = P^\mu \eta^{\nu\rho} - P^\nu \eta^{\mu\rho}, \\
\{M^{\mu\nu}, M^{\rho\sigma}\} = M^{\mu\sigma} \eta^{\nu\rho} + M^{\nu\rho} \eta^{\mu\sigma} - M^{\mu\rho} \eta^{\nu\sigma} - M^{\nu\sigma} \eta^{\mu\rho}.
\end{gather}

The action of Lorentz generators on the Fourier modes is given by
\begin{equation}
\{M^{\mu\nu}, a(\vec{k})\} = \eta^{\mu\mu'}\eta^{\nu\nu'}D_{\mu'\nu'} a(\vec{k}),
\end{equation}
where $D_{\mu\nu}$ is a realization of the Lorentz group in terms of the differential operators
 \begin{eqnarray}\label{realization}
D_{01}&=& {-}\sqrt{\vec{k}^2 + m^2}\  \partial_{k^1}\ \equiv iK_1,\label{mK1}\\
D_{02}&=& {-}\sqrt{\vec{k}^2 + m^2}\  \partial_{k^2}\ \equiv iK_2,\label{mK2}\\
D_{12}&=& k^1 \partial_{k^2} - k^2  \partial_{k^1} \equiv iJ. \label{mJ}
\end{eqnarray}

	One can check that the generators $J$, $K_1$ and $K_2$ obey the $SO(1,2)$ algebra
	\begin{equation}\label{SO12}
		[K_1,K_2] = - iJ,\quad [K_1,J] = -i K_2,\quad [K_2,J]=i K_1.
	\end{equation}

%With these generators, one can write variations for the field $\phi$ and its momentum $\pi$ in the Hamiltonian formalism as
%\begin{equation}
%\delta \phi = \epsilon_\mu \{\phi, P^\mu\}, \quad \delta \pi = \epsilon_\mu \{\pi, P^\mu\}.
%\end{equation}

%%%%%%%%%%%%%%%%%%%%%%%%%%%%%%%%%%%%%%%%%%%%%%%
\subsection{Supertranslations}\label{KS}
%%%%%%%%%%%%%%%%%%%%%%%%%%%%%%%%%%%%%%%%%%%%%%%
In order to construct a canonical realization of BMS$_3$ 
we follow the procedure of \cite{Longhi:1997zt} to construct supertranslations. 
The idea is to generalize the ordinary three dimensional momenta $k^\mu$ to an infinite set of ``supermomenta" and to generalize the realization of the charge of the translations on-shell (\ref{momenta}).

\subsubsection{Massive case}
Consider the $k_0>0$ sheet of the mass-shell hyperboloid representation of the hyperbolic plane $\mbox{H}_2$,
\begin{equation}\label{hyperboloid}
-k_0^2+k_1^2+k_2^2 =-m^2,
\end{equation} 
in a space with ambient Minkowski metric
\begin{equation}\label{ambient}
\dif s^2 = -\dif k_0^2 + \dif  k_1^2 + \dif k_2^2.
\end{equation}
The manifold  $\mbox{H}_2$ is invariant under the isometries of the metric, that is $ISO(1,2)$. We can parametrize (\ref{hyperboloid})  for $k_0>0$ as
\begin{eqnarray}\label{parametrization}
k_0 &=& m z,\label{k0}\\
k_1 &=& m \sqrt{z^2-1}\cos\phi,\label{k1}\\
k_2 &=& m  \sqrt{z^2-1}\sin\phi,\label{k2}
\end{eqnarray}
with $z\in[1,+\infty)$, $\phi\in[0,2\pi)$. Notice that $k_1$ and $k_2$ vanish at $z=1$. 
In these coordinates, the Lorentz generators (\ref{mK1}), (\ref{mK2}), (\ref{mJ}) are given by
\begin{eqnarray}
K_1 &=& -i\frac{z}{\sqrt{z^2-1}} \sin\phi \frac{\partial}{\partial\phi} + i \sqrt{z^2-1}\cos\phi \frac{\partial}{\partial z},\label{mmK1}\\	
K_2 &=& i\frac{z}{\sqrt{z^2-1}} \cos\phi \frac{\partial}{\partial\phi} + i \sqrt{z^2-1}\sin\phi \frac{\partial}{\partial z},\label{mmK2}\\
J &=& -i\frac{\partial}{\partial\phi}.\label{mmJ}
\end{eqnarray}
 
The metric induced on $\mbox{H}_2$, { which is an}
%turns out to be that of,  
Euclidean $\mbox{AdS}_2$ with AdS radius equal to $m$, is given, in this parametrization, by
\begin{equation}
\label{induced}
\dif s^2_{\text{induced}} = m^2 \frac{1}{z^2-1} \dif z^2 + m^2 (z^2-1)\dif\phi^2.
\end{equation}
The boundary is located at { $z\rightarrow\infty$, and it is an sphere $S^1$ with radius $m\sqrt{z^2-1}$ and 
metric
$$
\dif s^2|_{\text{boundary}}=\lim_{z\to\infty}\frac{1}{m^2 (z^2-1)} \dif s^2_{\text{induced}} = \dif\phi^2.
$$
}
The Laplace-Beltrami operator (Appendix B) is given by 
\begin{equation}
\label{laplacian}
\nabla^2 = \frac{1}{m^2}\left( (z^2-1)\frac{\partial^2}{\partial z^2} + 2z \frac{\partial}{\partial z} + \frac{1}{z^2-1}\frac{\partial^2}{\partial \phi^2}
\right),
\end{equation}
and it is proportional to a Casimir the $2+1$ Lorentz group,
\begin{equation}\label{LaplacianCasimir}
m^2\nabla^2 = -J^2 + K_1^2 + K_2^2. 	
\end{equation}
It is immediate to check that the three dimensional momenta are eigenfunctions of this operator,
\begin{equation}\label{eigen}
\nabla^2 k_\mu = \frac{2}{m^2} k_\mu, \quad \mu =0,1,2,
\end{equation}
{The numerical constant 2 is the dimension of the hyperboloid.}
%This gave the idea to the authors in \cite{Longhi:1997zt} 
To generalize the momenta to  infinite ``supermomenta" we look for general eigenvectors of the Laplacian with the eigenvalue $2/m^2$,
\begin{equation}\label{geneigen}
\left( {-}\nabla^2 + \frac{2}{m^2}
\right)\Phi(z,\phi) =0.
\end{equation}
 
We look for solutions of the form
\begin{equation}
\Phi(z,\phi) = e^{i\ell\phi}f(z),
\end{equation}
where $e^{i\ell\phi}$ are the eigenfunctions of  $S^1$.
The differential equation for $f(z)$ is
\begin{equation}\label{eqf}
(1-z^2)f'' -2z f' + \left( 2-\frac{\ell^2}{1-z^2}  \right)f=0,
\end{equation}
with general solution 
\begin{equation}
\label{solf}
f(z) = C_1\ (z-\ell)\left(   \frac{z+1}{z-1} \right)^{\dfrac{\ell}{2}} + C_2\  (z+\ell)\left(   \frac{z-1}{z+1} \right)^{\dfrac{\ell}{2}} 
\end{equation}
The first solution does not behave well at $z=1$ for $\ell>0$, while the second one is not well behaved for $\ell<0$, and one of the solutions becomes the other one by changing $\ell \leftrightarrow -\ell$.  Since the two dimensional momenta are regular for 
$z=1$, we are interested in the general solution to (\ref{geneigen}) that is regular at $z=1$, given by
 \begin{eqnarray}
w_\ell (z,\phi) &=&   e^{i\ell\phi} \left(   \frac{z-1}{z+1} \right)^{\dfrac{\ell}{2}} (z+\ell),\quad \ell \geq 0,\label{wsols1}\\
\widehat{w}_\ell (z,\phi) &=&  e^{i\ell\phi} \left(   \frac{z+1}{z-1} \right)^{\dfrac{\ell}{2}} (z-\ell),\quad \ell < 0,\label{wsols2}
\end{eqnarray}
which can also be written in a more compact form as
\begin{equation}\label{wsols3}
w_\ell (z,\phi) =  e^{i\ell\phi} \left(   \frac{z-1}{z+1} \right)^{\dfrac{|\ell|}{2}} (z+|\ell|),\quad \ell \in\Z.
\end{equation}
The functions $w_\ell (z,\phi)$ are the infinite set of ``supermomenta" that we are looking for.

	Notice that
	\begin{equation}\label{cba_asymp_massive}
	w_\ell(z,\phi)=e^{i\ell\phi}z + O(1/z),
	\end{equation}
	and we can define $w_\ell|_{\text{boundary}}(z,\phi)=e^{i\ell\phi}\, z$. This means that the generalized momenta have, for all $\ell$, the same asymptotic behavior as the ordinary momenta.

Alternatively, one can use the set of real functions
\begin{eqnarray}
u_\ell (z,\phi) &=&   \cos\ell\phi \left(   \frac{z-1}{z+1} \right)^{\dfrac{\ell}{2}} (z+\ell),\quad \ell \geq 0,\label{usols}\\
v_\ell (z,\phi) &=&   \sin\ell\phi \left(   \frac{z-1}{z+1} \right)^{\dfrac{\ell}{2}} (z+\ell),\quad \ell \ge 0.\label{vsols}
\end{eqnarray}

Notice that the three dimensional momenta can be written in terms of these functions as
\begin{eqnarray}
u_0(z,\phi) &=&   z =\frac{1}{m} k_0,\\
u_1(z,\phi) &=&   (z^2-1)^\frac{1}{2} \cos\phi = \frac{1}{m} k_1,\\
v_1(z,\phi) &=&   (z^2-1)^\frac{1}{2} \sin\phi = \frac{1}{m} k_2.
\end{eqnarray}

{ In terms of the $2$-dimensional momenta, the functions (\ref{wsols3}) can be written as
 \begin{eqnarray}
\omega_\ell (k_1,k_2) &=&
\left(\dfrac{k_1}{\sqrt{k_0^2-m^2}} + i\dfrac{k_2}{\sqrt{k_0^2-m^2}} \right)^{\ell} \\
\nonumber
& \cdot &\left(\dfrac{k_0 - m}{k_0+m}\right)^{|\ell|/2} \left(\frac{k_0}{m} + |\ell|\right).
 \end{eqnarray}
 This can be further simplified to yield
 \begin{eqnarray}
\omega_\ell (k_1,k_2) &=&
 \left(k_1+ik_2\right)^\ell |\vec{k}|^{-\ell} \left(\dfrac{\sqrt{m^2 + |\vec{k}|^2}-m}{\sqrt{m^2+|\vec{k}|^2} +m}\right)^{|\ell|/2} 
 \nonumber\\
 &\cdot& \left(\sqrt{1+|\vec{k}|^2/m^2} + |\ell|\right),\label{cba_wk_massive}
 \end{eqnarray}
 or 
 \begin{eqnarray}
 \omega_\ell (k_1,k_2) &=& \left(k_1+ik_2\right)^\ell \left(\sqrt{m^2+|\vec{k}|^2}+\text{sgn}(\ell) \,m\right)^\ell\nonumber\\ &\cdot& \left(\sqrt{1+|\vec{k}|^2/m^2} + |\ell|\right).
 \label{cba_wk_massive2}
 \end{eqnarray}
 
}

One can check that the subspace of functions spanned by $u_0$, $u_1$, $v_1$, or, alternatively, by $w_0$, $w_1$ and $w_{-1}$, is invariant under the action of the $2+1$ Lorentz group. 
In general the action of the Lorentz generators on the $w_\ell$ is given by
\begin{eqnarray}
K_1 w_\ell &= & -\frac{i}{2}   (\ell-1) w_{\ell +1} + \frac{i}{2}   (\ell+1) w_{\ell -1},\label{K1w}\\
K_2 w_\ell &= & -\frac{1}{2}   (\ell-1) w_{\ell +1} - \frac{1}{2}   (\ell+1) w_{\ell -1}, \label{K2w}\\ 
J w_\ell &=& \ell w_\ell.\label{Jw}
\end{eqnarray}
Defining $K_{\pm} = K_1 \pm i K_2$
one has
\begin{equation}
K_\pm w_\ell = i (1\mp \ell) w_{\ell\pm 1},\label{Kpmw}
\end{equation}
and therefore $K_\pm w_\ell$ are raising and lowering operators.

Since each function $w_\ell$ defines a (super)translation in the phase space of a massless scalar particle, we define the supertranslations generators as 
\begin{equation}\label{Pl}
P_\ell = \int \tilde{\dif k}\ w_\ell(\vec k)\ \bar a(\vec k)\,a(\vec k).
\end{equation} 
It is easy to check that these supertranslations commute
\begin{equation}\label{Pl1}
\left\{
P_\ell, P_{\ell'}
\right\} = 0. 
\end{equation} 
{
Their action on the Fourier modes is given by
\begin{equation}\label{supermomentaFourier}
	\left\{P_l, a(\vec k)\right\} = i w_\ell a(\vec k).
\end{equation}

%while on the boundary we have  
%\begin{equation}\label{supermomentaFourier}
%	\left\{P_l|_{\text{boundary}}, a(\vec k)\right\} = i r e^{i\ell\phi}a(\vec k).
%\end{equation} 
%

Let us see how Lorentz generators act on them. A Lorentz generator ${\cal O}$ is represented by
\begin{equation} \label{eq:Oina}
{\cal O}= \int \tilde{\dif k}\ \bar a(\vec k)\, O_{\vec k}\, a(\vec k),	
\end{equation} 
with $O_{\vec k}$ a first order differential operator in $\vec k$. For instance, for the rotation $J$ one has 
$$
O_{\vec k} = -i \left(k_1\frac{\partial}{\partial k_2} - k_2  \frac{\partial}{\partial k_1}        \right),  
$$
while for the boost generators $K_i$,
$$
O_{\vec k} =  i \sqrt{k_1^2+k_2^2+m^2}\frac{\partial}{\partial k_i} .
$$

One can show that
\begin{equation}
\left\{
{\cal O}, P_\ell
\right\} = - i \int \tilde{\dif k} a(\vec k) \bar a(\vec k)  O_{\vec k} w_\ell(\vec k)
\end{equation}
and hence it suffices to know the action of the generators on the funtions $w_\ell$. In particular, one gets
\begin{eqnarray}
\left\{  J, P_\ell \right\} &=& - i \ell P_\ell,\label{canJml}\\
\left\{  K_1, P_\ell \right\} &=& \frac{1}{2}  (1-\ell) P_{\ell +1}	+  \frac{1}{2} (1+\ell) P_{\ell -1}, \label{canK1ml}\\	
\left\{  K_2, P_\ell \right\} &=& -\frac{i}{2}  (1-\ell) P_{\ell +1}	+ \frac{i}{2}   (1+\ell) P_{\ell -1}, \label{canK2ml}	\\
\left\{  K_\pm, P_\ell \right\} &=&  (1\mp \ell) P_{\ell\pm 1}. \label{canKpml}
\end{eqnarray}
This is the analogous in 2+1 dimensions of the four dimensional BMS algebra \cite{BMS-1}  \cite{Longhi:1997zt}. The generalization of the algebra to include superrotations will be discussed after we consider the massless case.

 Relations (\ref{Pl1}) and (\ref{canJml})--(\ref{canKpml}) imply, at the quantum level, that  the Hilbert space of one-particle states supports a unitary irreducible representation of the Poincare group, and at the same time a unitary reducible representation of the BMS$_3$ group.  In contrast with the gravitational approach, our canonical realization of the supertranslations is unbroken.

\subsubsection{Massless case}
Now we want to construct the canonical realization of BMS associated  to a three
dimensional massless free scalar field.
In this case what previously was a hyperboloid is now a cone
\begin{equation}
-k_0^2+k_1^2+k_2^2 =0,
\end{equation}
that can be parametrized as 
\begin{eqnarray}
k_0 &=& r,\\
k_1 &=& r \cos\phi,\\
k_2 &=& r \sin\phi,
\end{eqnarray}
with  $r>0$ for $k_0>0$ and $\phi\in[0,2\pi)$. This can be obtained from (\ref{k0})---(\ref{k2}) putting $z=r/m$ and then letting $m\to 0$. \cite{Gomis:2015ata} However, a non-degenerate  induced metric does not exist and the standard construction of the Laplace operator fails. Instead,   we scale the operator in (\ref{geneigen}) by $m^2$ and replace $z=r/m$,
\begin{eqnarray}
\lefteqn{D\equiv -m^2 \Delta + 2}\\
& =& -\left(
\left\{\left( \frac{r}{m} \right)^2 -1  \right\} m^2 \partial_r^2 + 2 r \partial_r
+ \frac{1}{\left( \frac{r}{m}\right)^2 -1} \partial_\phi^2
\right) + 2.\nonumber
\end{eqnarray}
In the limit $m\to 0$ one gets
\begin{equation}\label{Dml}
D_\text{massless} = - r^2 \partial_r^2 - 2 r \partial_r +2,
\end{equation}
which turns out to be independent of $\phi$. 
As in the massive case, we look for solutions of
\begin{equation}
D_\text{massless}\Phi(r,\phi)=0
\end{equation}
 of the form
$$
\Phi(r,\phi)= e^{i\ell\phi} f(r).
$$

The function $f(r)$ must obey
\begin{equation}\label{fmassless}
-r^2 f'' -2 r f' + 2 f =0.
\end{equation}
This equation has independent solutions $f_1(r)=r$ and $f_2(r)= 1/r^2$, and hence the regular solution at $r=0$ is $f(r)=r$. The supermomenta that we are looking for are 
\begin{equation}\label{wmassless}
w_\ell(r,\phi) =  r\  e^{i\ell\phi},\quad \ell\in\Z,
\end{equation}
up to a normalization constant.
The expression in terms of momenta is given by
\begin{equation}
\omega_\ell(\vec{q}) = \dfrac{\left(q^1 + iq^2\right)^\ell}{\left((q^1)^2 + (q^2)^2\right)^{\frac{\ell-1}{2}}}.
\end{equation}
Notice that in the massless case the dependence of $r$ in the boundary and in the bulk is the same, {that is, there are no corrections in $1/r$ like in the massive case.}

The $SO(1,2)$ generators on the cone are given by
\begin{eqnarray}
J &=& - i \frac{\partial}{\partial\phi},\label{Jml}\\
K_1 &=& i r \cos\phi  \frac{\partial}{\partial r} - i \sin\phi \frac{\partial}{\partial\phi},\label{K1ml}\\
K_2 &=& i r \sin\phi  \frac{\partial}{\partial r} + i \cos\phi \frac{\partial}{\partial\phi}.\label{K2ml}
\end{eqnarray}

As in the massive case, it turns out that the purely differential part of $D_\text{massless}$ is proportional (actually equal in this case) to one of the 
{ two}
Casimirs of $SO(2,1)$,
\begin{equation}\label{LaplacianCasimirsMassless}
- r^2 \frac{\partial^2}{\partial r^2} - 2 r \frac{\partial}{\partial r} = -J^2 +K_1^2+ K_2^2.	
\end{equation}	

Notice that these generators, or more precisely $J$, $K_{\pm} = K_1 \pm iK_2$,  can be written as particular cases ($n=0,\pm1$) of the more general expression
\begin{equation}\label{cba_Ln_Lorentz}
L_n = e^{in\phi} \left(-\dfrac{\partial}{\partial \phi} + inr\dfrac{\partial}{\partial r}\right). 
\end{equation}
 The action of $SO(1,2)$ generators on the eigenfunctions $w_\ell$ is exactly the same found for the massive case
\begin{eqnarray}
J w_\ell &=& \ell w_\ell,\label{actJml}\\
K_1 w_\ell &=& -\frac{i}{2} (\ell-1)  w_{\ell +1} + \frac{i}{2} (\ell+1)  w_{\ell -1},\label{actK1ml}\\
K_2 w_\ell &=& -\frac{1}{2} (\ell-1)  w_{\ell +1} - \frac{1}{2} (\ell+1)  w_{\ell -1},\label{actK2ml}
\end{eqnarray}
which, in particular, shows that the subspace spanned by $w_0$, $w_1$ and $w_{-1}$ is invariant under  $SO(1,2)$.

The supertranslations are still given by 
\begin{equation}\label{P2}
P_\ell = \int \tilde{\dif k}\ w_\ell(\vec k)\ \bar a(\vec k)\,a(\vec k).
\end{equation} 
Together with the Lorentz generators, they constitute  a realization of the \BMSt algebra. {The action of supertranslations on the Fourier modes is analogous to the massive case. Induced  representations of
BMS$_3$ have been constructed in \cite{Barnich:2014kra}\cite{Barnich:2015uva}.

%%%%%%%%%%%%%%%%%%%%%%%%%%%%%%%%%%%%%%%%%%%%%%%
\subsection{Superrotations}\label{KS1}
%%%%%%%%%%%%%%%%%%%%%%%%%%%%%%%%%%%%%%%%%%%%%%%

We will construct here infinite families of operators generalizing the Lorentz algebra, for both the massless and massive cases.

\subsubsection{Massless case}
In the massless case, one can generalize \eqref{cba_Ln_Lorentz} for arbitrary $n\in\Z$ and write down
\begin{equation}
L_{n} = e^{in\phi}\left(-\dfrac{\partial}{\partial\phi} + inr\dfrac{\partial}{\partial r}\right), \quad n \in \Z .\label{cba_Ln}
\end{equation}
One can check that these $L_n$ obey also the Witt algebra
\begin{equation}\label{LnLm}
[L_n, L_m] = i(n-m) L_{n+m},
\end{equation}
and that 
{
\begin{equation}\label{Lnwl}
L_n w_\ell = i(n-\ell) w_{n+\ell}.
\end{equation}
}

In terms of $\vec{k}$, the differential operators (\ref{cba_Ln}) can be written as 
\begin{eqnarray}
L_{n} &=& \left(k_1^2+k_2^2\right)^{-n/2} \left(k_1 + ik_2\right)^{n} \left(\left\{ink_1 + k_2\right\}\dfrac{\partial}{\partial k_1}\right.\nonumber\\  &+&\left. \left\{ink_2 - k_1\right\}\dfrac{\partial}{\partial k_2}\right), \label{eq:Lnk}
\end{eqnarray}
and analogously to the case of supertranslations, we define the on-shell generators of superrotations as 
\begin{equation}\label{Rn_massless}
\mathcal{R}_{n} = \int \tilde{dk}\, \bar{a}(\vec{k})L_n a(\vec{k}),
\end{equation}
which, due to (\ref{LnLm}) and (\ref{Lnwl}), realize the  algebra 
\begin{eqnarray}
\{\mathcal{R}_m, \mathcal{R}_n\} &=& -i\int \tilde{dk}\, \bar{a}(\vec{k})[L_m, L_n] a(\vec{k})\nonumber\\ 
&=& (m-n)\mathcal{R}_{m+n},
 \\
\{\mathcal{R}_m,P_n\} &=& -i \int \tilde{dk}\, a(\vec{k})\bar{a}(\vec{k}) L_m w_n (\vec{k})\nonumber\\ 
&=& (m-n)P_{m+n},
\label{realizationLn_massless}\\
\{ P_n, P_m\} &=& 0.
\end{eqnarray}
This is the \BMSt algebra introduced in \cite{Barnich:2011ct}. 

{

	\subsubsection{Massive case}
	In the massive case, we do not have an initial guess for the form of the  superrotations. One may  notice, however, that Lorentz generators, once written in the form $\xi^\alpha \partial_\alpha,\,\,$ $\alpha=z,\phi$, satisfy the following equations
	\begin{equation} \label{eq:keyeq}
	D \xi^z = 0, \quad \nabla_\alpha \xi^\alpha = 0,
	\end{equation}
	where %the Lorentz generators have the form $\xi^\alpha \partial_\alpha$ and
	$D = -m^2 \Delta + { 2}$. Thus one can try, in the massive case, to generalize these operators by solving first $D \xi^z = 0$ and then computing $\xi^\phi$ from   $\nabla_z\xi^ z+\nabla_\phi \xi^\phi=0$.

	Clearly, 
	\begin{equation}
	\xi^z = e^{in\phi}\left(\dfrac{z-1}{z+1}\right)^{|n|/2} (|n|+z), \quad n\in \Z,
	\end{equation}
	since this is the solution of the PDE for massive supertranslations. From the divergence  equation, and  using that
	$\nabla_z\xi^ z+\nabla_\phi \xi^\phi=\partial_z\xi^ z+\partial_\phi \xi^\phi$ (see Appendix B),
	 one can integrate the angular term to obtain
	\begin{equation}
	\xi^\phi = -e^{in\phi} \left(\dfrac{z-1}{z+1}\right)^{|n|/2} \dfrac{(|n|(|n|+z)+z^2-1)}{in(z^2-1)} + f(z),
	\end{equation}
	with $f(z)$ an arbitrary function that we set to zero.
 
	Thus, one may try to define superrotation generators  as
	\begin{eqnarray}
	T_n &=& e^{in\phi} \left(\dfrac{z-1}{z+1}\right)^{|n|/2}\left(-\dfrac{|n|(|n|+z)+z^2-1}{z^2-1} \dfrac{\partial}{\partial\phi} \right.\nonumber\\
	&+& \left. in(|n|+z) \dfrac{\partial}{\partial z}\right), \quad n\in \mathbb{Z},
	\label{Tnmassive}
	\end{eqnarray}
	where we have multiplied all terms by a factor $in$. However, these operators do not form an algebra (this can be seen when computing  the commutator of $T_n$ with opposed sign indices, except in the case $n=\pm 1$).  Instead, we can define two infinite-dimensional set of generators, each containig the Lorentz part, according to
	\begin{eqnarray}
	{\cal L}_n  &=& T_n,\quad n\geq -1,
	\label{Lnright}\\
	{{\cal Q}}_n  &=& T_n,\quad n\leq 1.
		\label{Lnleft}
	\end{eqnarray}
	Both sets of differential operators satisfy the algebra
	\begin{eqnarray}
	\left[ {\cal L}_n, {\cal L}_m \right]&=& i(n-m){\cal L}_{n+m},\ n,m\geq -1,\label{wittright}\\
    \left[ {{\cal Q}}_n, {{\cal Q}}_m \right] &=& i(n-m){{\cal Q}}_{n+m},\ n,m \leq 1.\label{wittleft}
	\end{eqnarray}
	One has, for the lowest  values of $n$,
	\begin{equation}
	{\cal Q}_0={\cal L}_0 = -iJ, \ {\cal L}_1={\cal Q}_1 = K_{+},\ {\cal L}_{-1}={{\cal Q}}_1 = - K_{-}.
	\end{equation}
 Furthermore, the functions $w_n$ associated to each set provide a realization of the corresponding algebras,
 \begin{eqnarray}
 {\cal L}_n w_m &=& i(n-m)w_{n+m},\ n,m\geq -1,\label{rwittright}\\
 	{{\cal Q}}_n w_m&=& i(n-m)w_{n+m},\ n,m\leq 1.\label{rwittleft}
 	\end{eqnarray}
 Defining now the generators of superrotations as in (\ref{Rn_massless}) for each set one can construct 
  realizations of two subalgebras of the \BMSt algebra. To sum up, it is possible to extend the set of Lorentz generators to the right with the ${\cal L}_n$ and to the left with the ${{\cal Q}}_n$, but,  in contrast with what happens in the massless case, it is not possible to merge both extensions into a single algebra.

     The first equation  in (\ref{eq:keyeq}), which we obtained by generalizing the one satisfied by the Lorentz generators, is clearly non-covariant, but we will show next that, due to the geometry of the mass-shell manifold, it is, in fact, one of the components of a geometrical equation. 
     
     The Lorentz generators are the only solutions of the Killing equation
     \begin{equation}\label{killing}
     g^{\mu\alpha}\nabla_\alpha \xi^\nu + g^{\nu\alpha}\nabla_\alpha \xi^\mu =0.
     \end{equation}
     
     In order to generalize the generators of Lorentz transformations one could consider an equation of the form
     \begin{equation}\label{killing5}
     g^{\mu\alpha}\nabla_\alpha \xi^\nu + g^{\nu\alpha}\nabla_\alpha \xi^\mu = G^{\mu\nu}
     \end{equation}
     with $G$ symmetric and covariantly divergenceless,
     \begin{equation}\label{killing6}
     \nabla_\mu G^{\mu\nu} =0.
     \end{equation}
     Condition (\ref{killing6}) is instrumental for what we want to do. Notice, however, that we are not assuming that $G$ is proportional to the metric, and hence (\ref{killing5}) is different from the conformal Killing equation.
     
     We now take the covariant derivative $\nabla_\mu$ of (\ref{killing5}). Using that $\nabla_\alpha g^{\mu\nu}=0$ and (\ref{killing6}), imposing $\nabla_\mu \xi^\mu =0$, and using  $[ \nabla_\mu, \nabla_\alpha] \xi^\mu =  R^\mu_{\ \beta\mu\alpha} \xi^\beta  $, one arrives at
     \begin{equation}\label{campiglia1}
     g^{\mu\alpha}\nabla_\mu\nabla_\alpha \xi^\nu+ g^{\nu\alpha} R^\mu_{\ \beta\mu\alpha} \xi^\beta = 0.
     \end{equation}
     Using the explicit form of the components of the Riemann curvature tensor given in Appendix B, it turns out that
     \begin{equation}
     g^{\nu\alpha} R^\mu_{\ \beta\mu\alpha} \xi^\beta = -\frac{1}{m^2} \xi^\nu,
     \label{riemann1}
     \end{equation}
     and  (\ref{campiglia1}) boils down to \cite{ft0}
     \begin{equation}
     g^{\mu\alpha}\nabla_\mu\nabla_\alpha \xi^\nu = \frac{1}{m^2}\xi^\nu.
     \label{campiglia2}
     \end{equation}
     Evaluating (\ref{campiglia2}) for $\nu=z,\phi$ yields the pair of coupled equations
      \begin{eqnarray}
      \Delta_S \xi^z - \frac{2z}{m^2}(\partial_z\xi^z + \partial_\phi\xi^\phi) - \frac{2}{m^2}\xi^z	 &=& 0,\label{eqq1}\\
      \Delta_S \xi^\phi +\frac{2z}{m^2} \left(\partial_z\xi^\phi +\frac{1}{(z^2-1)^2} \partial_\phi \xi^z\right)&=&0,\label{eqq2}
      \end{eqnarray}
      where $\Delta_S$ is the scalar Beltrami-Laplace operator (\ref{laplacian}).
      However, due to 
      \begin{equation}\label{original1}
      0=\nabla_z \xi^z +\nabla_\phi \xi^\phi = \partial_z \xi^z + \partial_\phi \xi^\phi,
      \end{equation}
      equation (\ref{eqq1}) can be simplified to  
      \begin{equation}\label{original2}
      \Delta_S \xi^z  -\frac{2}{m^2} \xi^z = 0.
      \end{equation}
      Equations (\ref{original2}) and (\ref{original1}) were our starting point for constructing the superrotation generators, and now have received a sound geometrical foundation, \textit{i.e.} (\ref{campiglia1}). Furthermore,  and one can check that   (\ref{eqq2})  is satisfied  by the $\partial_\phi$ parts of ${\cal L}_n$ and ${{\cal Q}}_n$.
      
      Let us finally notice that the $1$-forms $l_n$ associated to ${\cal L}_n$ (and likewise for ${{\cal Q}}_n$),
      \begin{eqnarray}
      l_n &=&  m^2 e^{in\phi} \left(\dfrac{z-1}{z+1}\right)^{n/2}
      \left(i\frac{n(n+z)}{z^2-1} \dif z\right.\nonumber\\
      & &
      \left. -(n(n+z)+z^2-1)\dif\phi   \right), \quad n\in \mathbb{Z}.
      \end{eqnarray} 
      turn out to be eigenvectors of the Hodge-Laplace-de Rham operator $\tilde\Delta$,\cite{ft02}
      \begin{equation}\label{mainresult}
      \tilde \Delta l_n = -\frac{2}{m^2} l_n.
      \end{equation}
     This adds to the geometrical meaning of our construction, and could be useful for further generalizations.

}

\section{\label{NLBMS}Non-local BMS symmetries of the Klein-Gordon Lagrangian}

In this section we will prove that the KG action is invariant under supertranslations and superrotations, and construct the corresponding Noether charges.\cite{ft2} We will present explicit expressions only for the massless case.}

\subsection{Noether Charges of Supertranslations}

For the classical Klein-Gordon field, the Fourier modes can be written in terms of the fields $\phi$ and $\pi$ as
\begin{gather}
a(\vec{k})=\int d^2x\, e^{-ikx}\left(k^0 \phi(t,\vec{x}) + i\pi(t,\vec{x})\right), \label{eq:Fou1}\\
\bar{a}(\vec{k})=\int d^2x\, e^{ikx}\left(k^0 \phi(t,\vec{x}) - i\pi(t,\vec{x})\right), \label{eq:Fou2}
\end{gather}
where %$k^0 = \sqrt{\vec{k}^2 + m^2}$ in the massive case 
$k^0 = |\vec{k}|$.

When doing a supertranslation transformation on the fields using $P_\ell = \int \tilde{dk}\, \bar{a}(\vec{k})a(\vec{k}) \omega_\ell$ as the generator, one obtains
\begin{eqnarray}
\delta_{ST} \phi &=& \{\phi,\epsilon^\ell P_\ell\}\nonumber\\
 &=& \int \tilde{dk}\, (-i) \varepsilon^{\ell} \omega_{\ell} \left( a(\vec{k}) e^{ikx} - \bar{a}(\vec{k}) e^{-ikx}\right), \\
\delta_{ST} \pi &=& \{\pi, \varepsilon^{\ell} P_{\ell}\}\nonumber\\
 &=&  \int \tilde{dk} \,(-1) k^{0} \varepsilon^{\ell} \omega_\ell \left(a(\vec{k}) e^{ikx} + \bar{a}(\vec{k}) e^{-ikx}\right),
\end{eqnarray}
which can be written in terms of the fields using \eqref{eq:Fou1} and \eqref{eq:Fou2} as \cite{ft1}
\begin{gather}
\delta_{ST} \phi = \varepsilon^{\ell} \int d^2y \, \left[f_{\ell}(\vec{x}-\vec{y})\phi(t,\vec{y}) + g_{\ell}(\vec{x}-\vec{y})\pi(t,\vec{y})\right], 
\label{nonlocal}\\
\delta_{ST} \pi = \varepsilon^{\ell} \int d^2 y \, \left[h_{\ell} (\vec{x}-\vec{y}) \phi(t,\vec{y}) + f_{\ell} (\vec{x}-\vec{y}) \pi(t,\vec{y})\right],
\label{nonlocal2}
\end{gather}
where
\begin{align} 
&f_\ell(\vec{x}) = 2\int \tilde{dk} \,\omega_{\ell} (\vec{k}) k^0 \sin(\vec k\cdot \vec x), \label{eq:fl} \\
&g_\ell(\vec{x}) = 2\int \tilde{dk}\, \omega_{\ell} (\vec{k}) \cos(\vec k\cdot \vec x), \label{eq:gl} \\
&h_{\ell}(\vec{x}) = -2\int \tilde{dk}\, \omega_\ell (\vec{k}) {k^0}^2 \cos(\vec k\cdot \vec x). \label{eq:hl}
\end{align}
Notice the symmetry properties $f_\ell (-\vec{x}) = -f_\ell (\vec{x})$, $g_\ell (-\vec{x}) = g_\ell (\vec{x})$ and $h_\ell (-\vec{x}) = h_\ell (\vec{x})$, and that $\nabla^2 g_\ell (\vec{x}) = h_\ell (\vec{x})$.
 
Another important aspect to notice here concerns the values of $f_\ell$, $g_\ell$ and $h_\ell$ depending on the parity of $\ell$. One can check that for $\ell$ odd, $g_\ell = h_\ell = 0$, since their integrands are odd functions, and for $\ell$ even, $f_\ell = 0$, due to the same reason. This observation implies that a particular supertranslation will not use simultaneously information from a field and its momentum, but from only one of them. Thus, if $\ell$ is even, $\delta_{ST}\phi$ will depend only on the field momentum, whereas if $\ell$ is odd, $\delta_{ST} \phi$ will need just the value of the field itself.

Now we would like to see if we can extended the on shell symmetry to an off-shell Noether symmetry of the massless KG lagrangian. We consider the off-shell realization of (\ref{nonlocal})(\ref{nonlocal2}).
The variation of the lagrangian (\ref{action}) with $m=0$ under this transformation is
\begin{eqnarray}
\delta L &=&  \int\dif^2x\dif^2 y\ [
h(\vec x-\vec y)\phi(t,\vec y)\dot\phi(t,\vec x) \nonumber\\ &+& f(\vec x-\vec y)\pi(t,\vec y)\dot\phi(t,\vec x)\nonumber\\
&+& f(\vec x-\vec y)\dot\phi(t,\vec y)\pi(t,\vec x) + g(\vec x-\vec y)\dot\pi(t,\vec y)\pi(t,\vec x)\nonumber\\
&-& h(\vec x-\vec y)\phi(t,\vec y)\pi(t,\vec x) - f(\vec x-\vec y)\pi(t,\vec y)\pi(t,\vec x)\nonumber\\
&-& \vec\nabla_x f(\vec x-\vec y) \phi(t,\vec y)\cdot \vec\nabla\phi(t,\vec x)\nonumber\\ &-& \vec\nabla_x g(\vec x-\vec y)\pi(t,\vec y) \cdot \vec\nabla\phi(t,\vec x)
].\label{cba_ca5}	
\end{eqnarray}
The second and third terms cancel each other due to $f(-\vec x)=-f(\vec x)$, while the sixth and seventh terms (the latter upon using $\vec\nabla_x f(\vec x-\vec y)=-\vec\nabla_y f(\vec x-\vec y)$ and integration by parts with respect to $y$) cancel each one by themselves. Finally, the eighth term can be made to cancel the fifth one by integrating by parts the $\vec\nabla\phi(t,\vec x)$ and imposing
\begin{equation}
\nabla^2 g = h,\quad \text{and} \quad g(\vec x)=g(-\vec x),
\label{cba_ca6}
\end{equation}
(which implies also that $h(\vec x)=h(-\vec x)$).
One is left then with
\begin{eqnarray}
\delta L &=& 	\int\dif^2x\dif^2 y[
h(\vec x-\vec y)\phi(t,\vec y)\dot\phi(t,\vec x) \nonumber\\
&+& g(\vec x-\vec y)\dot\pi(t,\vec y)\pi(t,\vec x)
]\nonumber \\
&=& \dot F,
\label{cba_ca7}
\end{eqnarray}
with
\begin{eqnarray}
F &=& \frac{1}{2} \int\dif^2x\dif^2 y[
h(\vec x-\vec y)\phi(t,\vec y)\phi(t,\vec x)\nonumber\\ &+& g(\vec x-\vec y)\pi(t,\vec y)\pi(t,\vec x)
],
\label{cba_ca8}
\end{eqnarray}
and where $g(-\vec x)=g(\vec x)$ and $h(-\vec x)=h(\vec x)$ have also been used. The conserved charge is given by
\begin{equation}
Q = \int\dif^2 x\  \pi(t,\vec x)\delta\phi(t,\vec x) - F,
\label{cba_ca9}
\end{equation}
and one immediately gets
\begin{eqnarray}
Q &=& \int\dif^2 x\dif^2 y\ (
f(\vec x-\vec y)\pi(t,\vec x)\phi(t,\vec y)\nonumber\\  &+& \frac{1}{2}g(\vec x-\vec y)\pi(t,\vec x)\pi(t,\vec y) 
\nonumber\\
&-& \frac{1}{2} h(\vec x-\vec y)\phi(t,\vec y)\phi(t,\vec x)
),
\label{cba_ca10}	
\end{eqnarray}
which has the form of the canonical generators $P_\ell$ \eqref{Pl} (in terms of $\phi$ and $\pi$) for supertranslations.

\subsection{Noether Charges of Superrotations}

When acting over the Fourier modes, assuming they vanish sufficiently fast for high momentum (that is, boundary terms can be ignored), one obtains the simple transformation
\begin{equation}
\{\mathcal{R}_n, a(\vec{q})\} = i L_n a(\vec{q})
\end{equation}
\begin{equation}
\{\mathcal{R}_n, \bar{a}(\vec{q})\} = iL_n \bar{a}(\vec{q})
\end{equation}
Again, one can lift from on-shell to off-shell  the variations of the fields $\phi$ and $\pi$ in the Hamiltonian formalism under a superrotation
\begin{align}
\delta_{SR} \phi & = \int \tilde{dk} (\delta a(\vec{k}) e^{ikx} + \delta \bar{a}(\vec{k}) e^{-ikx}) \nonumber\\
& = \int d^2 y\, \left\{\phi(t,\vec{y})F_n (\vec{x},\vec{y}) + \pi(t,\vec{y}) G_n (\vec{x},\vec{y})\right\} \label{eq:dphiSR}\\
\delta_{SR} \pi & = \int \tilde{dk}(-ik^0) (\delta a(\vec{k}) e^{ikx} - \delta \bar{a}(\vec{k}) e^{-ikx}) \nonumber\\
& = \int d^2 y\, \left\{\phi(t,\vec{y}) \tilde{H}_n (\vec{x},\vec{y}) + \pi(t,\vec{y}) \tilde{I}_n (\vec{x},\vec{y})\right\} \label{eq:dpiSR}
\end{align}
where we have used the expressions of Fourier modes in terms of the field and momentum \eqref{eq:Fou1} and \eqref{eq:Fou2}  off-shell, where 
\begin{align}
&\lefteqn{F_n (\vec{x},\vec{y}) =}\nonumber\\
& -i\int \tilde{dk}\, \left[e^{ikx}(L_n e^{-iky}k^0) + e^{-ikx}(L_n e^{iky}k^0)\right] \\
&G_n (\vec{x},\vec{y}) = \int \tilde{dk}\, \left[e^{ikx}L_n e^{-iky} - e^{-ikx}L_n e^{iky}\right] \\
&\lefteqn{\tilde{H}_n (\vec{x},\vec{y}) =}\nonumber \\
& -\int \tilde{dk}\, k^0 \left[e^{ikx}(L_n e^{-iky}k^0) - e^{-ikx}(L_n e^{iky}k^0)\right] \\
&\tilde{I}_n (\vec{x},\vec{y}) = -i\int \tilde{dk}\,k^0 \left[e^{ikx}L_n e^{-iky} + e^{-ikx}L_n e^{iky}\right]
\end{align}
In contrast with the supertranslation case, the functions involved in the non-local transformation do not depend solely on the difference $\vec{y}-\vec{x}$ but on different combinations of these variables. More explicitly
\begin{eqnarray}
F_n (\vec{x},\vec{y}) &=& 
 -i\int \tilde{dk}\, 2\omega_n(\vec{k})
 \left[in\cos(\vec{k}(\vec{y}-\vec{x}))
 \right.\nonumber\\
 &-&\left.
  (in\vec{y}\cdot\vec{k} + \vec{y}\times\vec{k})\sin(\vec{k}(\vec{y}-\vec{x}))\right] \label{eq:Fn} \\
G_n (\vec{x},\vec{y}) &=& i\int \tilde{dk}\, 2\frac{\omega_n(\vec{k})}{k^0}
 \nonumber\\
& &
\left[(in\vec{y}\cdot\vec{k} + \vec{y}\times\vec{k})\cos(\vec{k}(\vec{y}-\vec{x}))\right] \label{eq:Gn} \\
\tilde{H}_n (\vec{x},\vec{y}) &=& i\int \tilde{dk}\, 2k^0 \omega_n(\vec{k})\left[i n \sin(\vec{k}(\vec{y}-\vec{x}))
\right.\nonumber\\
&+&\left.
(in\vec{y}\cdot\vec{k} + \vec{y}\times\vec{k})\cos(\vec{k}(\vec{y}-\vec{x}))\right] \\
\tilde{I}_n (\vec{x},\vec{y}) &=& i\int \tilde{dk}\, 2\omega_n(\vec{k})
 \nonumber\\
& &
\left[(in\vec{y}\cdot\vec{k} + \vec{y}\times\vec{k})\sin(\vec{k}(\vec{y}-\vec{x}))\right]
\end{eqnarray}
where $\vec{y}\times\vec{k}\equiv y_1k_2-y_2k_1$. Here, there is an important remark to make concerning the parity of $|n|$: if $|n|$ is odd $F_n=\tilde{I}_n=0$, and if $|n|$ is even $G_n=\tilde{H}_n=0$. For the case of rotations, $L_0 = -iJ$;
\begin{align}
&F_0 (\vec{x},\vec{y}) = 2i\int \tilde{dk}\, k^0 \left[\vec{y}\times\vec{k}\right]\sin(\vec{k}(\vec{y}-\vec{x})) \\
&G_0 (\vec{x},\vec{y}) = 2i\int \tilde{dk}\, \left[\vec{y}\times\vec{k}\right] \cos(\vec{k}(\vec{y}-\vec{x}))
\end{align}
By symmetry properties $G_0(\vec{x},\vec{y})=0$, and one can substitute $F_0$ in \eqref{eq:dphiSR} and show that the usual rotation is recovered:
\begin{equation}
\delta_{SR_0} \phi = i\left(x_1\partial_{x_2} \phi(t,\vec{x}) - x_2 \partial_{x_1} \phi(t,\vec{x})\right).
\end{equation}
{Recall that when we defined superrotations in section \ref{KS1}, we used the truncated (time-independent) form of Lorentz generators constructed in (\ref{eq:redefM}). Thus, when trying to recover ordinary boosts, which involve time, we will need to redefine superrotations to take this under consideration. The final form for superrotations will be
\begin{equation}
\delta_{\mbox{ordinary SR}_n} \phi = -nt\, \delta_{ST_n} \phi + \delta_{SR_n} \phi,
\end{equation}
which now accounts for time translations. With this definition, one can check that for $n=1$, a combination of ordinary boosts is recovered
\begin{eqnarray}
\delta_{\mbox{ordinary SR}_1} \phi &=& t \partial_{x_1} \phi(t,\vec{x}) + i t\partial_{x_2} \phi(t,\vec{x})
\nonumber \\ & &
 - x_1 \pi(t,\vec{x}) - i x_2 \pi(t,\vec{x}).
\end{eqnarray}
Hence, the true superrotations generators will be a combination of the already constructed ones plus a proportional term depending on supertranslations. This can be written as follows
\begin{equation}
\mathcal{G}_n = -n t P_n + \mathcal{R}_n.
\label{eq:gt}
\end{equation}
}

 The generators $\mathcal{R}_n$ can be written off-shell as
%  follows
%\begin{align}
%\mathcal{R}_n =&\, i\int \tilde{dk}\,d^2x\,d^2y\,e^{in\phi}e^{i\vec{k}(\vec{y}-\vec{x})} \left[(k^0)^2\phi(t,\vec{x})\phi(t,\vec{y})\left(in\vec{y}\cdot\vec{k} + \vec{y}\times\vec{k}+n\right)\right. \nonumber\\
%	&\left. + ik^0\phi(t,\vec{x})\pi(t,\vec{y})\left(in\vec{y}\cdot\vec{k} + \vec{y}\times\vec{k}\right) - ik^0\pi(t,\vec{x})\phi(t,\vec{y}) \left(in\vec{y}\cdot\vec{k} + \vec{y}\times\vec{k}+n\right)\right. \\
%	&\left. + \pi(t,\vec{x})\pi(t,\vec{y})\left(in\vec{y}\cdot\vec{k} + \vec{y}\times\vec{k}\right) \right] \nonumber
%\end{align}
%which can be simplified defining $\tilde{F}_n$ and $\tilde{G}_n$ as the functions $F_n$ and $G_n$ in \eqref{eq:Fn} and \eqref{eq:Gn}, respectively, but with an extra $k^0$ under the integral sign. Analogously, $H_n$ and $I_n$ are defined as the corresponding functions but for a missing $k^0$. With all this,
\begin{eqnarray}
\mathcal{R}_n &=& \dfrac{1}{2}\int d^2y\, d^2x\, [\phi(t,\vec{x})\phi(t,\vec{y})(\tilde{H}_n + i\tilde{F}_n)
\nonumber\\
& &  - i\phi(t,\vec{x})\pi(t,\vec{y}) (\tilde{G}_n - i \tilde{I}_n) \nonumber \\
& & -i \pi(t,\vec{x}) \phi(t,\vec{y}) \left(H_n + i F_n\right) \nonumber\\
& & + \pi(t,\vec{x}) \pi(t,\vec{y}) \left(G_n - i I_n\right) ].
\end{eqnarray}
where $\tilde{F}_n$ and $\tilde{G}_n$ as the functions $F_n$ and $G_n$ in \eqref{eq:Fn} and \eqref{eq:Gn}, respectively, but with an extra $k^0$ factor under the integral sign, while  $H_n$ and $I_n$ are defined as the corresponding functions but with an additional  $1/k^0$ factor.  

%Furthermore, one can check that $\mathcal{R}_n$ operators do not transform in general solutions into solutions (for $n=0$, it does), which indicates that they do not generate a conserved charge. {This is so because, as noticed above, these operators only recover the truncated Lorentz generators. That is why we defined $\mathcal{G}_n$,} which is indeed a conserved charge, as can be deduced by computing its temporal derivative (notice $H=P_0$)
The $\mathcal{G}_n$ given in (\ref{eq:gt}) are constants of motion
\begin{eqnarray}
\dfrac{d\mathcal{G}_n}{dt} &=& \partial_t \mathcal{G}_n + \{\mathcal{G}_n, H\}\nonumber\\
 &=& - n P_n - nt\{P_n, P_0\} + \{\mathcal{R}_n, P_0\}\nonumber\\&=& - nP_n + nP_n = 0,
\end{eqnarray}
where we have used that $H=P_0$.
This was expected, since the Lagrangian is invariant under Lorentz transformations.

The new field variations, $\delta_{\mathcal{G}_n} \phi = - nt\delta_{ST} \phi + \delta_{SR} \phi$, are solutions on-shell of the massless Klein-Gordon equation:
\begin{align}
\Box \delta_{\mathcal{G}_n} \phi &= - nt\Box \delta_{ST} \phi + n\partial_t \delta_{ST} \phi + \Box \delta_{SR} \phi\nonumber\\ 
& = n\partial_t \delta_{ST} \phi + \Box \delta_{SR} \phi \nonumber\\
&= n \epsilon^n \int d^2 y \, \left[f_n (\vec{x}-\vec{y}) \dot{\phi}(t,\vec{y}) + g_n (\vec{x}-\vec{y}) \dot{\pi}(t,\vec{y})\right] \nonumber\\
& \quad+ \epsilon^n \int d^2 y\, \left[\left\{\nabla_{\vec{x}}^2 F_n(\vec{x},\vec{y}) - \nabla_{\vec{y}}^2 F_n(\vec{x},\vec{y}) \right\} \phi(t,\vec{y}) \right.
\nonumber \\
&\quad\left.+ \left\{\nabla_{\vec{x}}^2 G_n(\vec{x},\vec{y}) - \nabla_{\vec{y}}^2 G_n(\vec{x},\vec{y}) \right\} \pi(t,\vec{y})\right]. 
\end{align}
Using now the on-shell condition $\dot{\pi} = \ddot{\phi} = \nabla^2 \phi$, integrating by parts and using symmetry properties of $g$ in the first integral, and expanding the second one, and using then $h_n = \nabla^2 g_n$,  it is immediate to see that $\Box\delta_{\mathcal{G}_n} \phi =0$.
 The algebra of charges is
\begin{equation}
\{\mathcal{G}_n, \mathcal{G}_m\} = (n - m) \mathcal{G}_{n+m}.
\end{equation}
Thus we have found another realization of superrotations, which now reduce to the true Lorentz generators as defined in \eqref{eq:TrueL}. Indeed,
\begin{align}
&\mathcal{G}_0 = \mathcal{R}_0 = -i\int \tilde{dk}\, \bar{a}(\vec{k}) J a(\vec{k}) = M^{12}, \\
&\mathcal{G}_1 = -t P_1 + \mathcal{R}_1 = -t \int \tilde{dk}\, \bar{a}(\vec{k}) (k_1 + i k_2) a(\vec{k}) \nonumber\\
&+ \int \tilde{dk}\, \bar{a}(\vec{k}) K_{+} a(\vec{k}) = -M^{01} - iM^{02}, \\
&\mathcal{G}_{-1} = t P_{-1} + \mathcal{R}_{-1} = -t \int \tilde{dk}\, \bar{a}(\vec{k}) (k_1 - i k_2) a(\vec{k})\nonumber\\
& - \int \tilde{dk}\, \bar{a}(\vec{k}) K_{-} a(\vec{k}) = -M^{01} + iM^{02}.
\end{align}

\section{Conclusions and Outlook}

{
Using the canonical formalism for a real scalar field, a realization of the BMS group in 3 dimensions has been constructed in the space of Fourier modes, for both the massive and the massless case. In the massless case, the superrotation extension of this group can also be constructed by generalizing the Lorentz group in a similar way as it is done for supertranslations. 

{ In the massive case,  we have constructed in an heuristic way
a set of generators which generalize those of the Lorentz group and reproduce the corresponding part of the \BMSt algebra. We have shown how our starting equations arise in a geometrical setup, and have also obtained an equation for the $1$-forms associated to the generators. 
 
However, unlike what happens in the massless case, the superrotation generators must be split into two different extensions of the Lorentz algebra, each spanning a subalgebra of \BMSt.
%Since this splitting is particular of the $2$ spatial dimensions, we conjecture that a similar  mechanism will not exist in higher dimensions, and that superrotations cannot be defined in  our canonical framework for $(d+1)$-spacetimes with $d>2$.

}

At the quantum level, the Hilbert space of one-particle states supports a unitary irreducible representation of the Poincare group, and at the same time a unitary reducible representation of the BMS$_3$ group. Both are realized in an unbroken way.

The BMS$_3$ transformations are realized as symmetries of the KG action in terms of linear non-local functionals of the field and the canonical momentum. The corresponding  conserved Noether charges have been computed.

Besides obtaining a better understanding of the extension of superrotations in the massive case, some further questions are still open for future work.
 
There was the belief, in the gravitational approach, that BMS was not present in higher dimensions. From the viewpoint considered in this paper, there is no reason to think so, and the method presented here could help to investigate it. In fact, it can be proved that the canonical realization of BMS in higher dimensions does exist \cite{future}.
	
One could also try to add a fermionic field to the present model in order to get a field supersymmetric theory, and see whether  there are  still conserved charges generated by the extended BMS transformations. In the gravitational approach this has been studied in \cite{Awada:1985by}\cite{Barnich:2014cwa}\cite{Barnich:2015sca}\cite{Lodato:2016alv}.
	
{	
Finally, there is the question of the physical interpretation of the BMS symmetries and charges in the framework that we have used. A possible way to throw light into this issue is to try to construct particle models exhibiting these symmetries, using the method of nonlinear realizations.\cite{Gomis:2006xw}
}	
{
	We also conjecture that the non-locality of the transformations is due to the fact that they are computed for fields depending only on the standard space-time coordinates, and that they would become local for fields depending also on the \textit{supercoordinates} associated to the \textit{supermomenta}, \textit{i.e.} the generators of supertranslations.  
}	
 
}
%%%%%%%%%%%%%%%%%%%%%%%%%%%%%%%%%%%%%%%%%%%%%%%
\section*{Acknowledgements}\addcontentsline{toc}{section}{Acknowledgements}
We acknowledge interesting discussions with  Glenn Barnich, Jacques Distler, Willy Fischler, Marc Henneaux, Giorgio Longhi, Sonia Paban, Alfonso Ramallo, Jorge Russo and Steven Weinberg. We also thank H.~A.~Gonz\'{a}lez and H.~Afshar for pointing out their work to us, and Jan Govaerts and Diego Delmastro for comments on the realization of superrotations in the massive case.

JG is grateful to Steven Weinberg for the hospitality and support during a visit to the Physics Dept. Theory Group of the Univ. of Texas at Austin,  where this 
paper was completed. This material  is  based upon work supported in part  by the National Science Foundation under Grant Number PHY-1620610 and with support from 
The Robert A. Welch Foundation, Grant No. F-0014. 

JG  has been supported  by FPA2013-46570-C2-1-P, 2014-SGR-104 (Generalitat de Cata\-lunya) and Consolider CPAN and by
the Spanish goverment (MINECO/FEDER) under project MDM-2014-0369 of ICCUB (Unidad de Excelencia Mar\'\i a de Maeztu).
CB is partially supported by  the Generalitat de Catalunya through project 2014 SGR 267 and by the Spanish government (MINECO/FEDER) under project CICYT DPI2015-69286-C3-2-R.

%%%%%%%%%%%%%%%%%%%%%%%%%%%%%%%%%%%%%%%%%%%%%%%

\begin{appendix}
%%%%%%%%%%%%%%%%%%%%%%%%%%%%%%%%%%%%%%%%%%%%%%%%

%%%%%%%%%%%%%%%%%%%%%%%%%%%%%%%%%%%%%%%%%%%%%%%%

\section{Behavior of $f_\ell$, $g_\ell$ and $h_\ell$}
In this appendix we study in some detail the functions which appear in the non-local transformations constructed in Section 3. We only consider the massless case, although similar, but more complicated, expressions can be obtained in the massive case using (\ref{cba_wk_massive}). 

The functions $f_\ell(\vec x)$ can be written as
$$
f_\ell(\vec x) = \frac{1}{(2\pi)^2} \int \dif^2 k\  \omega_\ell (\vec k) \sin(\vec k\cdot\vec x),
$$
where (in the massless case)
$$
\omega_\ell (\vec k) = |\vec k| e^{i\ell\phi_k},
$$
with
$$
\cos\phi_k = \frac{k_1}{|\vec k|},\quad \sin\phi_k = \frac{k_2}{|\vec k|}.
$$
Notice that $\omega_{-\ell}(\vec k) = \omega_{\ell}^*(\vec k)$.

The Fourier transform of $f_\ell$ is
$$
\hat{f}_\ell(\vec q) = \int\dif^2 x \  f_\ell(\vec x) e^{-i\vec q\cdot \vec x}  = \frac{1}{2i}\left(\omega_\ell(\vec q) - \omega_\ell(-\vec q)  \right).
$$
Since $\phi_{-q}=\phi_q + \pi$ one has that
\begin{equation}
\omega_\ell(-\vec q) = \omega_\ell(\vec q) e^{i\ell\pi} = (-1)^\ell \omega_\ell(\vec q)
\end{equation}
and
\begin{equation}
\hat{f}_\ell(\vec q) = \begin{cases} 0 & \text{if $\ell$ even},\\ -i \omega_\ell(\vec q) & \text{if $\ell$ odd}.
\end{cases}
\end{equation}

For a general value of $\ell$ one can write $\omega_\ell$ as a function of $q^1$ and $q^2$ as
\begin{equation}
\omega_\ell(\vec{q}) = \dfrac{\left(q^1 + iq^2\right)^\ell}{\left((q^1)^2 + (q^2)^2\right)^{\frac{\ell-1}{2}}}.
\end{equation}
\vglue1mm
Hence, for $\ell$ odd,  $\hat{f}_\ell(\vec q)$ grows as $|\vec q|$, while $\lim_{|\vec q|\to 0}  \hat{f}_\ell(\vec q) =0$.

In particular, for $\ell=1$ one has that $\hat f_1 (\vec q) = -i (q_1+iq_2)$. The physical components are
\begin{eqnarray*}
	\frac{\hat{f}_1 (\vec q)+ \hat{f}_{-1}(\vec q)}{2} &=& -i q_1,\\
	\frac{\hat{f}_1 (\vec q)- \hat{f}_{-1}(\vec q)}{2i} &=& -i q_2,
\end{eqnarray*}
which correspond to ordinary translations along the coordinate axes.
For $\ell=2m+1$, $m\geq 1$ one gets in the denominator of $\omega_\ell$ positive integer powers of $q_1^2+q_2^2=|\vec q|^2$, which implies that the transformation is non-local. For instance, for $\ell=3$,
\begin{eqnarray*}
	\hat{f}_3(\vec q) & =&   -i \frac{q_1^3 +3i q_1^2 q_2- 3 q_2^2 q_1 - i q_2^3}{q_1^2+q_2^2}.
\end{eqnarray*}

%\textcolor{blue}{
Similarly, one can see that
\begin{equation}
\hat{g}_\ell(\vec q) = \begin{cases} \frac{1}{|\vec{q}|} \omega_\ell(\vec q) , & \text{if $\ell$ even},\\ 0, & \text{if $\ell$ odd}.
\end{cases}
\end{equation}
For $\ell$ even one has, explicitly,
\begin{equation}
\hat{g}_\ell(\vec q) = \frac{(q_1+iq_2)^\ell}{(q_1^2+q_2^2)^{\ell/2}}.
\end{equation}
In particular, for $\ell=0$ one gets $\hat g_0(\vec q) =1$, that is, $g_0(\vec x)=\delta(\vec x)$, which yields a local transformation, corresponding to translations in time, while for $\ell\geq 2$ the transformations are non-local.

Finally, since $h_\ell(\vec x)= \nabla^2 g_\ell(\vec x)$, one has $\hat h_\ell(\vec q) = - |\vec q|^2 \hat g_\ell(\vec q)$ and
\begin{equation}
\hat{h}_\ell(\vec q) = \begin{cases} -|\vec{q}| \omega_\ell(\vec q) , & \text{if $\ell$ even},\\ 0, & \text{if $\ell$ odd}.
\end{cases}
\end{equation}
For $\ell$ even this is
\begin{equation}
\hat{h}_\ell(\vec q) = -\frac{(q_1+iq_2)^\ell}{(q_1^2+q_2^2)^{\ell/2-1}}.
\end{equation}
This yields local transformations for $\ell=0,2$ and non-local for $\ell\geq 4$.
%}

{
\section{Geometry of the mass-shell hyperboloid in (2+1)dimensions}
We list here some results about  of the geometry of the mass-shell hyperboloid of a massive particle in (2+1)dimensions that are useful for the construction of supertranslations and superrotations.

The metrics in $(z,\phi)$ coordinates (\ref{induced}) and its inverse are, in matrix form,  
\begin{equation}
g=\left(
\begin{array}{cc} \frac{m^2}{z^2-1} & 0 \\  0 & m^2 (z^2-1)\end{array}
\right),
\quad
g^{-1}=\left(
\begin{array}{cc} \frac{z^2-1}{m^2} & 0 \\  0 & \frac{1}{m^2 (z^2-1)}\end{array}
\right).
\end{equation}
The non-zero Christoffel symbols are
\begin{equation}
\Gamma^z_{zz} = - \frac{z}{z^2-1}, \quad   \Gamma^z_{\phi\phi} = -z (z^2-1), \quad  \Gamma^\phi_{z\phi}=\Gamma^\phi_{\phi z}=\frac{z}{z^2-1}.
\end{equation}
Given a vector field $\xi$ on the manifold,
\begin{equation}
\xi = \xi^z \partial_z + \xi^\phi \partial_\phi,
\end{equation}
we can construct an associated  $1$-form  using $g$,
\begin{equation}
\omega_\xi = \frac{m^2}{z^2-1}\xi^z \dif z + m^2 (z^2-1)\xi^\phi \dif\phi.
\end{equation}

The divergence of a vector field is
\begin{eqnarray}
\nabla_z \xi^z + \nabla_\phi \xi^\phi &=& \partial_z\xi^z + \Gamma^z_{z\alpha}\xi^\alpha + \partial_\phi \xi^\phi + \Gamma^\phi_{\phi\alpha}\xi^\alpha \nonumber \\
&=& \partial_z \xi^z - \frac{z}{z^2-1}\xi^z + \partial_\phi \xi^\phi + \frac{z}{z^2-1}\xi^z \nonumber \\
&=&  \partial_z \xi^z + \partial_\phi \xi^\phi,
\end{eqnarray}
so it coincides with the flat divergence.
The Beltrami-Laplace operator acting on  a function $f(z,\phi)$  is
\begin{eqnarray}
\Delta f &=&  \frac{1}{\sqrt{|g|}} \partial_\alpha \left(
\sqrt{|g|} g^{\alpha\beta}\partial_\beta f
\right)\nonumber\\ & =&  \partial_\alpha g^{\alpha\beta}\partial_\beta f + g^{\alpha\beta}\partial_\alpha\partial_\beta f
\nonumber \\
&=&   \frac{2z}{m^2}\partial_z f + \frac{z^2-1}{m^2}\partial_z^2 f + \frac{1}{m^2 (z^2-1)}\partial_\phi^2 f.
\label{escalarL} 
\end{eqnarray}
If we denote this scalar laplacian by $\Delta_S$,  on vector fields one has
\begin{eqnarray}
g^{\mu\alpha}\nabla_\mu \nabla_\alpha \xi^\nu
&=&
\Delta_S \xi^\nu + g^{\mu\alpha}\nabla_\mu (\Gamma^\nu_{\alpha\beta}\xi^\beta).
\end{eqnarray} 
 
\vglue1mm

The non-zero components of the Riemann curvature tensor are
\begin{eqnarray}
R^{z}_{\ \phi z \phi} &=& - (z^2-1), \  R^{z}_{\ \phi \phi z} = z^2-1,\nonumber\\  
 R^{\phi}_{\ z z \phi} &=& \frac{1}{z^2-1},\  R^{\phi}_{\ z\phi z} = - \frac{1}{z^2-1},
\end{eqnarray}
 and the Ricci scalar curvature is 
 \begin{equation}
 R = -\frac{2}{m^2}.
 \end{equation}

}

\end{appendix}

%%%%%%%%%%%%%%%%%%%%%%%%%%%%%%%%%%%%%%%%%%%%%%%


\begin{thebibliography}{99}
\bibitem{ft02}
The action of the Hodge-Laplace-de Rham operator on forms is related to that of the Laplace-Beltrami, or Bochner, laplacian $\Delta_B = g^{\alpha\beta} \nabla_\alpha\nabla_\beta$ by means of Weitzenb\"{o}ck identity.
\end{thebibliography}
\end{document}